\documentclass{llncs}

\usepackage{times}
\usepackage{helvet}
\usepackage{courier}
\usepackage{latexsym}
\usepackage{amsmath}
\usepackage{amssymb}
\usepackage{url}
\usepackage{ifpdf}
\usepackage{enumerate}
\usepackage{paralist}
\usepackage{alltt}
\usepackage{verbatim} 
\usepackage{color}
\usepackage[T1]{fontenc}
\usepackage[utf8]{inputenc}
\usepackage[english]{babel}
\usepackage{graphicx}

\def\qed{\hfill{}\ensuremath{\Box}}
   {\par\noindent{\bf Proof of #1}.\quad #2}%
   {\qed}
   {\par\noindent{\bf Proof of #1} (\emph{Sketch\/}).\quad #2}%
   {\qed}

\newlength{\phantomlength}

\def\ba{\begin{array}}
\def\ea{\end{array}}

\def\bitem{\vspace{0ex}\begin{itemize}}
\def\eitem{\end{itemize}\vspace{0ex}}

\newcommand{\egc}[0]{e.g.\ }

\begin{document}
\pagenumbering{gobble}
%
\title{CONDENSER: A Graph-Based Approach\\ for Detecting Botnets}
\author{Pedro Camelo\inst{1}\inst{2} \and Jo\~ao Moura\inst{1}\inst{2}\thanks{The work of J. Moura is supported by grant SFRH/BD/69006/2010 from Funda\c{c}\~ao para a Ci\^encia e Tecnologia (FCT) from the Portuguese Minist\'erio do Ensino e da Ci\^encia.} \and Ludwig Krippahl\inst{2}}
\institute{
AnubisNetworks R\&D, Amadora, Portugal\\
\texttt{\{pedro.camelo,joao.moura\}@anubisnetworks.com}
\and
CENTRIA, Universidade NOVA de Lisboa, Portugal\\
\texttt{pmcamelo@gmail.com}\;\; \texttt{joaomoura@yahoo.com}\;\;  \texttt{ludi@fct.unl.pt} 
}


\newenvironment{myprogram}[2]{{\begin{figure}\label{#1}}\par}{\vspace{-.3in}{\end{figure}}}

\pagestyle{plain}
\setcounter{secnumdepth}{2} 
\setcounter{tocdepth}{2} 

\maketitle

\def\mathrlap{\mathpalette\mathrlapinternal}%
\def\mathrlapinternal#1#2{\rlap{$\mathsurround=0pt#1{#2}$}}%

\makeatletter
\newdimen\@mydimen%
\newdimen\@myHeightOfBar%
\settoheight{\@myHeightOfBar}{$|$}%
\newcommand{\SetScaleFactor}[1]{%
    \settoheight{\@mydimen}{#1}%
    \pgfmathsetmacro{\scaleFactor}{\@mydimen/\@myHeightOfBar}%
}%

\newcommand*{\Scale}[2][3]{\scalebox{#1}{\ensuremath{#2}}}%

\newcommand*{\nct}[3]{%
    \SetScaleFactor{\vphantom{\ensuremath{#1#2}}}
    #1%
\mathrel{\Scale[\scaleFactor]{|\mathrlap{\kern-0.48ex\sim}\hphantom{\kern-0.41ex\sim}}_#3}%
    #2%
}%

\begin{abstract}
Botnets represent a global problem and are responsible for causing large financial and operational damage to their victims. They are implemented with evasion in mind, and aim at hiding their architecture and authors, making them difficult to detect in general. These kinds of networks are mainly used for identity theft, virtual extortion, spam campaigns and malware dissemination. Botnets have a great potential in warfare and terrorist activities, making it of utmost importance to take action against. 

We present CONDENSER, a method for identifying data generated by botnet activity. We start by selecting appropriate the features from several data feeds, namely DNS non-existent domain responses and live communication packages directed to command and control servers that we previously sinkholed. By using machine learning algorithms and a graph based representation of data, then allows one to identify botnet activity, helps identifying anomalous traffic, quickly detect new botnets and improve activities of tracking known botnets. 

Our main contributions are threefold: first, the use of a machine learning classifier for classifying domain names as being generated by domain generation algorithms (DGA); second, a clustering algorithm using the set of selected features that groups network communication with similar patterns; third, a graph based knowledge representation framework where we store processed data, allowing us to perform queries.
\end{abstract}

\section{Introduction} 
\label{section:intro}

Currently there is an large number of infections caused by malware present in all sorts of electronic devices. It is estimated that about 16\% to 25\%~\cite{AMLJS09} of Internet traffic in the world comes from communication between various types of malware. Moreover, because of the pervasive nature of devices such as smartphones and tablets which are increasingly replacing mobile phones and computers~\cite{Sma}, the number of infections is increasing and becoming more complex. More computational power, more tools as well as more opportunities and targets are available to malware develop, as they start exploring broader features within each device. 


Botnets are generally deployed for illegal purposes and are used for identity theft, massive spamming, distributed denial of service (DDoS),  government and industrial espionage and unauthorized use of computational resources \egc for bitcoin mining. Statistics estimate that roughly $75\%$~\cite{Trustwave2013} of all global email traffic is coming from spam, and that most of this traffic originates from botnets. Recent develops show us that such networks also have large potential in scenarios of warfare and terrorism, which is another reason to put an effort into stoping the authors that produce malware, and consequently to botnets. There are various governmental authorities, companies and organizations that struggle daily against these cybercriminals.\\

Identifying botnets can be done by malware analysis in controlled environments (\egc sandboxing or virtualization), or through analysis of malware communication with the Command and Control (C\&C) servers and other infected devices. The two main ways to detect botnets can be broadly divided into: Passive and Active. Passive detection is based on communication analysis of packets between the C\&C and infected devices, and has the advantage of providing no warning of the impending discovery of botnet infrastructure. Active sensing involves, in addition, a direct interaction with the botnet through the injection of network packets (\egc dns query to the domain nameserver). Botnets often implement evasion techniques, making it difficult to take action against. Some examples are, encrypted communication~\cite{Zhao2013}, domain generation algorithms~\cite{Schiavoni2013} (DGA), fast-flux and double fast-flux~\cite{Hogben2011}\footnote{Fast-flux is a DNS technique used by botnets in order to hide not only phishing activities but also sites delivering malware behind networks of compromised hosts that change very often and that act as proxies.}.




\subsection{Objectives}
Monitoring network traffic generates large volumes of data that must be processed in order to obtain useful information. In addition, botnet evasion techniques and the diversity of services that generate network traffic, which is usually encrypted, make it necessary to examine patterns on sets of features in order to detect botnet activity. The problem is, thus, one of inferring useful information from a large flux of complex data. In this paper we present a solution to this problem, using a combination of expert knowledge preprocessing followed by classification and clustering with machine learning algorithms.\\

This document is divided into four sections. In Section~\ref{section:intro}, we introduced the motivation, the context and gave an introduction to the topic of botnets. In Section~\ref{section:related} we discuss related work  and overview existing research within the topic as well as current solutions to the problem described and present and overview of relevant machine learning techniques. In Section~\ref{section:contributions} we present and evaluate the machine learning techniques we employed and present also a novel method for representing botnet communications as well as to characterise these botnets in terms of singly connected components of domains and IPs. We finish with conclusions and a discussion of future work in Section~\ref{section:conclusions}.

\section{Related Work}
\label{section:related}

There is growing interest in botnet research, both in the industry and in academia, motivated by the need to counter the criminal activity associated with botnets. In addition to general security considerations, a major issue in botnet research is the development of automated detection methods.


This research focused both on active and passive detection. For active detection we used DNS records analysis. For passive detection we identified three sub-categories: domain names analysis, flow analysis and packet analysis. For the purpose of this work we did not consider flow analysis since we were positioned in the middle of devices communication and NAT and IP pooling makes it harder to detect relevant IP changes.

The literature reports a diversity of approaches to traffic analysis, and thus, illicit activity detection. For example, Amini, et al.~\cite{Amini2006}, used Neural Networks to detect network intrusions; Ruehrup, et al.~\cite{Ruehrup2013} used common destinations as a way to compute connection sub-graphs to identify malicious P2P networks; Stalmans, et al. ~\cite{Stalmans} showed that geographic information of DNS records helped identify suspicious traffic using Morgan Indexes and the Geary Coefficient. Antonakakis, et al.~\cite{Antonakakis2012a} studied the behavior of some malware that use Domain Generation Algorithms to reach their C\&Cs, and Davuth, et al.~\cite{Davuth2013} showed that support vector machines could identify generated domain names with better performance than other classifiers, such as \textit{Na\"{i}ve Bayes} and \textit{C5.0} classification trees. Schiavoni, et al.~\cite{Schiavoni2013} concluded that domain name analysis alone is not sufficient to detect botnets, so they included IP and DNS analysis as a similarity feature to group same DGA domains. This diversity of methods, some using expert knowledge and others machine learning, inspired our combined approach.

We present next an overview of the machine learning techniques we use in our work.

\subsection{Classification: Support Vector Machines}
Machine learning is a "field of study that gives computers the ability to learn without being explicitly programmed", as defined by Arthur Samuel in 1959. In the field of cyber security and threat intelligence we highlight applications in the areas of malware detection, malicious traffic detection and spam detection.

In machine learning classification, support vector machines (SVM) are models created with supervised learning, associated with learning algorithms that analyze data and recognize patterns. They are mostly used for classification and regression analysis. In the case of binary SVMs, given a set of training examples each marked as belonging to one of two categories, an SVM training algorithm builds a model that assigns new examples into one category or the other, making it a non-probabilistic linear classifier. An SVM model is a representation of the examples as points in space, mapped so that the examples of the categories are divided as best as possible. Examples for classification are then mapped into that same space and predicted to belong to a category based on which side of the division they fall on.

In general, an SVM constructs a hyperplane or set of hyperplanes in a high or infinite dimensionality space. Intuitively, a good separation is achieved by the hyperplane that has the largest distance to the nearest training data point of any class (so-called functional margin). It is thus the case that, the larger the margin the lower the generalization error of the classifier.

\subsection{Clustering: Self-Organizing Maps}
A self-organizing map (SOM) is a type of artificial neural network that is trained using unsupervised learning, with the purpose of reducing the dimensionality of features into a low-dimensional representation of the input space. These self-organizing maps differ from traditional artificial neural networks in the sense that they use a neighborhood function to preserve the topological properties of the input space.


\subsection{Cluster Interpretation and Validation: Indexes}

The Davies-Bouldin~\cite{Davies1979} index -- represented as $DB_{k}$ -- identifies \textit{clusters} that are compact and distant from each other, in Equation~\ref{eq:dbdiam} the diameter for \textit{cluster} $c_{i}$ is obtained, where $n_{i}$ represents the number of points belonging to \textit{cluster} $c_{i}$. Symbol $z_{i}$ corresponds to the centroid of \textit{cluster} $c_{i}$ and $x$ is a point belonging to \textit{cluster} $c_{i}$.

\begin{equation}
  \label{eq:dbdiam}
  DB_{k} = \frac{1}{k}\sum\limits_{i=1}^k Max_{j=1,...,k,i\neq j} \bigg\{\frac{diam(c_{i})+diam(c_{j})}{eucl(c_{i} - c_{j})}\bigg\}  
\end{equation}

In $DB_{k}$, $k$ corresponds to total number of existing \textit{clusters}, $c_{i}$ and $c_{j}$ correspond to to the centroid of \textit{clusters} $i$ and $j$ respectively.

\begin{equation}
  \label{eq:dunndiss}
  diss(c_{i},c_{j}) = min_{x\in c_{i},y\in c_{j}}||x - y||
\end{equation}

Equation~\ref{eq:dunndiss} corresponds to the dissimilarity index between clusters $c_{i}$ and $c_{j}$, Equation~\ref{eq:dunndiam} is the diameter of the cluster $C$.

\begin{equation}
  \label{eq:dunndiam}
  diam(C) = max_{x,y\in C}||x - y||
\end{equation}

The Silhouette~\cite{L.Kaufman1990} index -- represented as $SI_{k}$ -- identifies the average membership of each point to all $k$ clusters, where $n$ is the total number of existing samples in the dataset, $a_{i}$ is the average distance between point $i$ and all points of its cluster; $b_{i}$ is the minimum average dissamilarity between point $i$ and all the formed clusters. In this metric, the partition with bigger $SI$ value is considered optimal.

\begin{equation}
  \label{eq:silh}
  SI_{k} = \frac{1}{n}\sum\limits_{i=1}^n\frac{(b_{i} - a_{i})}{max(a_{i},b_{i})}
\end{equation}

\section{Contributions}
\label{section:contributions}

Currently implemented modules include one Support Vector Machine (SVM) classifier for detecting suspicious domain names (generated by DGA); a Self-Organizing Map for data clustering; one from two IP reputation services; one from Maltracker\footnote{\url{https://maltracker.net/}}, a malware analysis platform that extracts information and network behavior of malware samples ran in sandboxed environments; one from an AnubisNetworks proprietary API informing whether an IP address had contacted one of AnubisNetworks' sinkholes; and finally we developed a module that crawls over DNS servers gathering historic information about domain names and IPs, with the goal of detecting common name servers and IP addresses used as C\&Cs, botnet domain rotation and botnets that operate using fast-flux and double fast-flux schemes. 

For the DGA classifier and as benign dataset, we used the Top 10,000 domains from Alexa, and as anomalous dataset we used a dataset of approximately 10,000 DGA domains provided by AnubisNetworks. We also automated the generation of such classifier modules by implementing a system for feature selection, training and evaluation of classifiers, making it simple not only to adapt the current DGA classifier to changes in the data stream, but also to include other classifiers in the future. The Self-Organizing Map groups data with similar patterns discarding the source and destination of the data communication since we want to find similar traffic that can have no intersection in its communications at all (\egc same botnet family, different campaign and operators).

\subsection{Graph Based Detection}
\label{section:graph}

To improve botnet detection, we created a modular application that extracts and aggregates packet data, by combining expert knowledge and machine learning classification and clustering. This system is composed by a set of modules that obtain information from external services and incoming packet data. The modular implementation makes it easier to adapt the system to different data streams, and the combination of expert knowledge rules and machine learning takes advantage, of both, known indicators of malicious activity and the capacity of machine learning algorithms to identify patterns in complex data. The output is a connected graph of the observed communications, including the data obtained from the modules, where nodes are IPs, domains, SOM Clusters representing communication with similar patterns and objects such as links to executable files and/or their signatures that were seen communicating with common destinations/IPs. With this system, we correlate information gathered from many sources, to detect suspicious sub-graph topologies and relations that allows us to characterise botnets in terms of singly connected components of domains and IPs.

\subsection{Discussion}
\label{section:discussion}

The presented work aims to identify all types of known botnet topologies (star, distributed, hybrid), by correlating information from various sources. We based our work in the existing related work, and we are able to identify common evasion techniques such as, domain names generated by domain generation algorithms (DGA), with the corresponding machine learning classifier. The precision of this classifier indicate a result of 77,9\% when ran against a set of domains generated by malware without a pre-defined dictionary in its implementation (\egc english names).

With our DNS crawler we gather information about normal and suspect domains, normally associated with low TTLs and high record changing (associated with fast-flux and double fast-flux). However there are some legitimate uses that can be miss-interpreted as these evasion methods. For example, Content Delivery Networks (CDNs) rely on low TTL records in their domains. This behavior can raise false positives if we only take the TTL value in consideration. For this reason we must take other information when we decide that we are over a new botnet family. Since our work relies in network communication, and taking in consideration malware that use the network as a way to communicate with the C\&C\footnote{Contrary to this, computer worms sometimes do not rely in any type of communication from a C\&C, or can became dormant for a long period of time relying in a physical way of spreading, like an usb stick. For these cases, we are not able to get data.}, we can gather information about all the malware lifecycle. 

The clustering algorithm provides a way to detect communication with similar patterns of known clusters even if the packet source and destination do not intersect, this characteristic provide a way to detect same malware families. When we consider the full system we have a way to detect not only the typical communication cluster, with enriched data, but also communication that matches already known communication patterns. With the information obtained from our sinkholes we are able to group communication that matches known botnet families.

Traffic that resulted in non-existent domain (NX-Domain) replies can not be compared with destinations that were "alive" when we take DNS and Geographic IP information in consideration, for this reason we didn't include features that match these information in our machine learning algorithms.

\subsection{Evaluation}
\label{section:results}

For the domain name classifier, we obtained $77,9\%$ of precision, using a set of 11 features. For the clustering algorithm we obtained a rating of $9.32$ with the Davies-Bouldin Index and of aproximately $0.27$ for the Silhouette Index with a total of $81$ features.

During our research we were not able to compare our results with work available in the literature because existing work only takes in consideration traffic inside universities campus, or inside small networks sometimes even with artificially generated traffic which is also referred in~\cite{Sommer2010}, and since we are positioned between internet operators, we are unable to extract some of the features proposed in related work.

\section{Conclusions and Future Work}
\label{section:conclusions}

We developed a system that aims to correlate information from various sources, including an automatic classifier taking domain names as inputs and deciding whether they are part of a DGA or not, information about live samples and sinkholed domains, and a clustering tool using Self-Organizing Maps to group network traffic having the same patterns. For checking if a domain name was generated by a DGA, our classifier has $77,9\%$ of precision, for the clustering algorithm used we obtained a rating of approximately $9.32$ using the Davies-Bouldin Index. As an auxiliary mechanism we developed a system that crawls over DNS servers, gathering historic DNS information which allows us to discover both fast-flux and double fast-flux domains as well as rotation of C\&C  servers. We store all this data in a graph that allows us to single out botnets as singly connected components of malicious servers (domains) and infected machines (IPs). 

As future work, we will proceed with our research by using continuous training algorithms for the clustering algorithms, such as implementing a growing and hierarchical extension of self-organizing maps, termed Growing Hierarchical Self-Organizing Maps~\cite{Rauber2002} (GHSOM).

\bibliography{botconf}
\bibliographystyle{plain}

\pagebreak

\appendix
\section{Considered Features}

\begin{table}[ht]\scriptsize
\caption{Domain Name Features}
\centering
\begin{tabular}{|l|l|p{6cm}|}
  \hline
  \textbf{Feature} & \textbf{Type} & \textbf{Feature Description} \\
  \hline \hline
  \texttt{consonantRatio} & Numeric & Consonant Ratio \\
  \texttt{consonantVowelRatio} & Numeric & Consonant Vowel Ratio \\
  \texttt{domainLength} & Numeric & Domain Length normalized as RFC 3986 (255 characters)\\
  \texttt{othersRatio} & Numeric & Other characters ratio \\
  \texttt{vocalRatio} & Numeric & Vowel Ratio \\    
  \texttt{digitRatio} & Numeric & Digit Ratio \\    
  \texttt{numRepeatsByUniGram} & Numeric & Number of string repetitions by Uni-gram analysis \\    
  \texttt{numRepeatsByBiGram} & Numeric & Number of string repetitions by Bi-gram analysis \\    
  \texttt{numRepeatsByTriGram} & Numeric & Number of string repetitions by Tri-gram analysis \\    
  \texttt{numRepeatsByTetraGram} & Numeric & Number of string repetitions by Tri-gram analysis \\    
  \texttt{lowFrequenceOccurrence} & Categoric & Domains starting and ending with a digit \\
  \hline
\end{tabular}
\label{table:ComplexityResults}
\end{table}

\begin{table}[h!]\scriptsize
\caption{URI Features}
\centering
\begin{tabular}{|p{5cm}|l|p{5cm}|}
  \hline
  \textbf{Feature} & \textbf{Type} & \textbf{Feature Description} \\
  \hline \hline
  \texttt{queryLength} & Numeric & URI Query Length \\
    \texttt{queryArgumentSize} & Numeric & Number of URI Query Arguments \\
    \texttt{uriPathLength} & Numeric & URI Path Length \\
    \texttt{uriPathLevelLength} & Numeric & URI Path Level Length \\
    \texttt{uriPathPlusLength} & Numeric & URI Path + Following Components Length \\
    \texttt{uriExistence} & Categoric & URI Path + Following Components Existence \\    
    \texttt{exe, bat, cmd, msi, com, drv, js, css, dat, ppt, doc, docx, txt, rtf, php, cgi, asp, aspx, html, xhtml, jsf, dll, png, jpg, bmp, bin, dll, zip, rar, swf, scr, wpad, pac, ini} & Categoric & URI Extension \\
    \texttt{unknownExtension} & Categoric & Unknown Extension (not in the feature list above) \\
    \texttt{unavailableExtension} & Categoric & Extension not available \\
    \texttt{consonantRatio} & Numeric & URI Base Consonant Ratio \\
    \texttt{vocalRatio} & Numeric & URI Base Vowel Ratio \\
    \texttt{consonantVowelRatio} & Numeric & URI Base Consonant Vowel Ratio \\
    \texttt{extensionLength} & Numeric & URI Extension Length \\        
  \hline
\end{tabular}
\label{table:ComplexityResults}
\end{table}

\begin{table}[h!]\scriptsize
\caption{Meta Features}
\centering
\begin{tabular}{|l|l|p{6.5cm}|}
  \hline
  \textbf{Feature} & \textbf{Type} & \textbf{Feature Description} \\
  \hline \hline
    \texttt{packetSize} & Numeric & TCP/IP Packet Size \\
    \texttt{packetSizeInexistence} & Categoric & Unknown TCP/IP Packet Size \\
    \texttt{200, 301, 400, 404, 413} & Categoric & Código de Resposta HTTP \\
    \texttt{unknownReplyCode} & Categoric & Unknown HTTP Reply Code (not present in the above feature list) \\
    \texttt{inexistentHttpRCode} & Categoric & HTTP Reply Code not available \\
    \texttt{HTTP/1.0, HTTP/1.1} & Categoric & HTTP Version \\
    \texttt{unknownHttpVersion} & Categoric & Unknown HTTP Version \\    
    \texttt{inexistentHttpVersion} & Categoric & HTTP Version not available\\        
  \hline
\end{tabular}
\label{table:ComplexityResults}
\end{table}

\end{document}